# Influence of strong electron irradiation on fluctuation conductivity and pseudogap in YBa$_2$Cu$_3$O$_{7-\delta}$ single crystals

A. L. Solovjov ●,[1,2,*] K. Rogacki ●,[2] N. V. Shytov ●,[1] E. V. Petrenko ●,[1] L. V. Bludova,[1] A. Chroneos ●,[3,4] and R. V. Vovk[5]

[1]*B. Verkin Institute for Low Temperature Physics and Engineering of the National Academy of Sciences of Ukraine, Kharkiv 61103, Ukraine*
[2]*Institute for Low Temperatures and Structure Research, Polish Academy of Sciences, Wroclaw 50-422, Poland*
[3]*Department of Materials, Imperial College, London SW7 2AZ, United Kingdom*
[4]*Department of Electrical and Computer Engineering, University of Thessaly, Volos 38221, Greece*
[5]*Department of Physics, V. N. Karazin Kharkiv National University, Kharkiv 61022, Ukraine*

The effect of high-energy electron irradiation on the temperature dependences of the resistivity $\rho(T)$, fluctuation conductivity (FLC), and pseudogap (PG) $\Delta^*(T)$ of YBa$_2$Cu$_3$O$_{7-\delta}$ (YBCO) single crystals containing virtually no twins was studied. A linear increase in the resistivity and a linear decrease in the superconducting (SC) transition temperature $T_c$ with increasing irradiation doses $\varphi$ were observed. For relatively small $\varphi$, the linear $T_c$ can be described by the Abrikosov-Gorkov (AG) pair breaking theory, and for large $\varphi$, by the Emery-Kivelson (EK) theory, which takes into account the suppression of $T_c$ by quantum phase fluctuations caused by irradiation defects. As FLC shows, at the average value of $\varphi_3 = 2.5 \times 10^{19}$ e/cm$^2$, which corresponds to the AG-EK crossover, the distance between the conducting CuO$_2$ planes, $d_{01}$, as well as the coherence lengths along the $c$ axis, $\xi_c(0)$, and the region of SC fluctuations, $\Delta T_{fl}$, increase sharply, and the two-dimensional contribution of the Maki-Thompson fluctuations (2D-MT) unexpectedly changes to the two-dimensional contribution of the Aslamazov-Larkin (2D-AL). Surprisingly, no features in $\rho(\varphi)$ and $T_c(\varphi)$ indicating the AG-EK crossover are observed. At the same time, at $\varphi_3$, a sharp increase in the opening temperature of PG, $T^*$, as well as the value of PG, $\Delta^*$, is observed, which indicates a possible decrease in DOS under the influence of defects. With a further increase in $\varphi$, all the parameters of PG and its dimensions are greatly reduced, and an unusual shape of $\Delta^*(T)$ is found. However, quite unexpectedly, at $\varphi_5 = 5.6 \times 10^{19}$ e/cm$^2$ the temperature dependences of both FLC and PG demonstrate curves typical for well-structured YBCO, regardless of the number of defects.

## I. INTRODUCTION

How SC pairing in high-temperature superconductors (HTSCs) enables the formation of Cooper pairs at temperatures well above 100 K [1,2] remains the most fundamental question in modern superconductivity physics. Understanding the SC pairing mechanism is of paramount importance for finding new HTSCs capable of room-temperature superconductivity. Unfortunately, this mechanism still remains uncertain. Studying another unusual property of HTSCs, known as the pseudogap, which opens in underdoped cuprate HTSCs of the REBCO type (RE = Y, Ho, Gd) at a characteristic temperature $T^* \gg T_c$ [1,3,4], should shed light on the problem [5] (and references therein). In cuprates, the temperature dependence of the resistivity $\rho(T)$ in the normal state is always linear up to very high temperatures, but deviates from linearity towards a decrease at $T < T^*$ [3,6]. It has been convincingly shown that below $T^*$ not only $\rho(T)$ decreases more than linearly, but also the density of states at the Fermi level (DOS) begins to gradually decrease [7,8], what is identified as the opening of a pseudogap [9]. As a result, numerous anomalies of electronic properties are observed in the PG state, associated with a decrease in the density of single-particle excitations and anisotropic rearrangement of the spectral density of charge carriers, most likely due to the rearrangement of the Fermi surface [1–4,10]. However, the physics of this process remain highly questionable [5].

Some researchers believe that pseudogap is due to SC phase fluctuations [11,12], antiferromagnetic spin fluctuations [1,6,13], spin density waves (SDWs) [14,15], charge density waves (CDWs) [16,17], charge ordering (CO) [3,10], and even pair density waves [18,19], which have nothing to do with paired fermions above $T_c$. However, the very variety of proposed models raises certain doubts regarding their conclusions [20]. Moreover, none of these models explains the observed deviation of $\rho(T)$ from linearity at $T \leqslant T^*$ and, more importantly, is unable to explain the absence of PG in the overdoped regime of cuprates, despite the fact that $T_c$ is still very high. In turn, another part of the researchers believes that PG appears as a result of the formation of paired fermions just below $T^*$, the so-called local pairs (LPs), or preformed Cooper pairs [4,5,21–25] and, thus, is a precursor of the transition of HTSCs to the SC state [26,27]. We share the second approach to the formation of PG and believe that LPs appear in cuprates as so-called strongly bound bosons (SBBs) at $T \leqslant T^*$, which

---

obey the Bose-Einstein condensation (BEC) theory. With decreasing $T$, LPs change their properties and near $T_c$ behave as incoherent Cooper pairs, which obey the BCS theory. Thus, at a certain temperature $T_{pair}$ in the interval $T^* > T_{pair} > T_c$ the BEC-BCS crossover should occur [5,28]. This approach allows us to explain most of the experimental facts observed in HTSCs [5], but it is still not generally accepted.

Another important issue concerning HTSCs is the role of defects. Single crystals of HTSC cuprates without impurities may contain defects in the form of twins and twin boundaries that arise during the "tetra-ortho transition" in the manufacturing process in order to minimize the elastic energy of the crystal [29,30]. YBCO single crystals may also contain point defects (oxygen vacancies), which is associated with nonstoichiometric oxygen content [31] (and references therein). Doping of YBCO with various impurities, which is currently widely studied in order to discover new properties of HTSC, usually leads to the appearance of additional defects in samples [32,33]. As a rule, various structural defects significantly affect the behavior of HTSCs, increasing the resistivity $\rho(T)$, but decreasing the superconducting transition temperature $T_c$ and increasing the width $\delta T_c$ of the SC transition [34–36]. However, these defects arise uncontrollably and are distributed randomly.

For that reason, the possibility of creating defects in a controlled manner is of considerable interest [35–37] (and references therein). This can be done, for example, by irradiating a HTSC sample with high-energy electrons. Such electrons may effectively shift any atoms of the crystal, forming numerous defects on different sublattices, the number of which can be controlled by choosing the irradiation dose, and, what is important, without changing the composition of the irradiated sample [31,37–43]. Obviously, irradiation defects will effectively affect not only the resistance, but also the excess conductivity, so the PG. Therefore, by studying the FLC and PG properties, one can obtain additional information both on the mechanism of local pairs formation above $T_c$ and on the mechanism of SC pairing in HTSC. This research is also of some practical interest, since HTSC devices are sometimes forced to operate under electron irradiation conditions [31] (and references therein).

Against the background of a huge number of works on HTSCs, the number of works devoted to the effect of electron irradiation on HTSC is extremely small [31,37–43] (and references therein). An attempt to study the effect of electron irradiation on the FLC and PG in YBCO single crystals was undertaken in our previous work [31]. However, the irradiation doses were comparatively small, and, most importantly, it was proven that our single crystals contained multiple defects in the form of twin boundaries. This led to some unusual resistive properties of the sample, differing from the available literature data [37–43], in particular, to a violation of the Matthiessen rule, and, apparently, to a specific behavior of the FLC and PG [31].

In this paper we present results of resistivity $\rho(T)$, fluctuation conductivity $\sigma'(T)$, and pseudogap $\Delta^*(T)$ measurements performed on optimally doped untwined YBa$_2$Cu$_3$O$_{7-\delta}$ (YBCO) single crystals, irradiated at low temperatures by high doses of electrons with energy 2.5 MeV. This made it possible to study the effect of irradiation defects on the properties of HTSCs in a wide range of irradiation doses increasing from $\varphi_1 = 0$ to $\varphi_{12} = 23 \times 10^{19}$ e/cm$^2$. One of the more remarkable results of this paper is a number of interesting but somewhat unexpected results obtained from the FLC and PG analysis, which we discuss in detail.

## II. EXPERIMENT

The YBCO single crystals were grown in Y$_2$O$_3$ crucibles by the conventional flux method, as described elsewhere [44]. Before detwinning, the crystals were subsequently annealed in air at different temperatures and quenched to room temperature to adjust their oxygen content. The samples were always quenched to room temperature at the end of the annealing to avoid long-range oxygen ordering phenomena, which often complicate the physics of YBCO [45]. Detwinning was performed at temperatures below 220 °C under a uniaxial pressure of ∼0.1 GPa [44]. Thus, all the YBCO samples prepared in this way are believed to be untwined single crystals. The single crystals with oxygen index $(7-\delta)$ of ≅ 6.96, the transition temperature $T_c$ in the range of 90–92 K, and a transition width $\delta T_c < 1$ K were studied. To estimate the oxygen content in our crystals, we compared the $T_c$ values, taken at $\rho(T) = 0$ from the extrapolated superconducting transition, with literature data [46]. Selected crystals were small-sized platelets with typical dimensions of $(1.5–2) \times (0.2–0.3) \times (0.01–0.06)$ mm$^3$, where the smallest size corresponded to the $c$ axis. The small size of the samples was necessary to minimize the irradiation time at a limited irradiation current. The $ab$ plane resistivity, $\rho_{ab}(T)$, was measured by the standard four-probe ac technique [47]. To ensure uniform current distribution in the central region of the sample, where voltage probes in the form of parallel strips were located at a distance of ∼1 mm between them, the current contacts were placed along opposite ends of the crystal and glued with silver epoxy. The contact resistances did not exceed 1 Ω. Temperature measurements were performed with a Pt sensor with an accuracy of about 1 mK [47].

The samples were immersed in liquid He, and an electron flux limited to $10^{14}$ (e/cm$^2$)/s was used to avoid heating the samples [48]. The irradiations were performed with 2.5 MeV electrons in the low-temperature facility of the Van de Graaff accelerator at the National Science Center of the Kharkiv Institute of Physics and Technology (Kharkiv, Ukraine). To ensure a uniform distribution of induced defects throughout the volume of the sample, the thickness of the samples (≈10–60 μm) was chosen to be significantly less than the electron penetration depth. The energy loss of 2.5 MeV electrons in YBCO can be estimated to be ∼1 keV/μm (see [38,39] and references therein) and does not exceed 3%–8% for the used crystal thickness. A specially designed helium cryostat [48] made it possible to measure the resistance after irradiation in the temperature range of 10 K < $T$ < 300 K.

## III. RESULTS AND DISCUSSION

### A. Resistance and critical temperature

The set of resistivity vs temperature results $\rho(T) = \rho_{ab}(T)$ of the optimally doped untwined single crystal YBa$_2$Cu$_3$O$_{7-\delta}$ (YBCO) with $T_c = 92.5$ K $(\varphi_1 = 0)$ and oxygen index

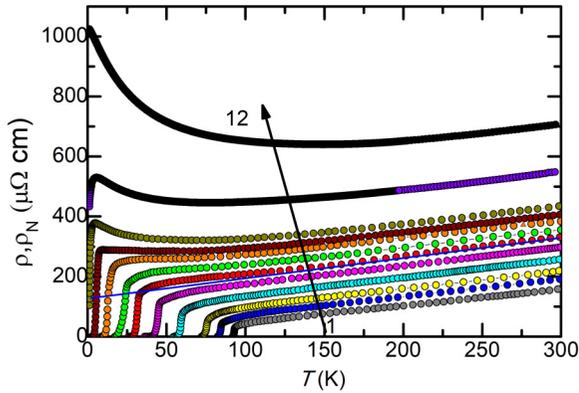

FIG. 1. Temperature dependences of the resistivity of an optimally doped YBa$_2$Cu$_3$O$_{7-\delta}$ single crystal $(7-\delta \cong 6.96)$ at different irradiation doses from $\varphi_1 = 0$ to 1.3, 2.5, 3.9, 5.6, 6.9, 7.9, 8.7, 10.0, 11.2, 15.5 and $\varphi_{12} = 23 \times 10^{19}$ e/cm$^2$. Blue line denotes $\rho_N(T)$ at $\varphi_6 = 6.9 \times 10^{19}$ e/cm$^2$, above which first signs of a metal-insulator transition appear and PG disappears.

$(7-\delta) \sim 6.96$, measured for irradiation doses from $\varphi_1 = 0$ to $\varphi_{12} = 23 \times 10^{19}$ e/cm$^2$, is shown in Fig. 1. As expected, with an increase in the irradiation dose $\varphi$, $T_c$ rapidly decreases to extrapolated $T_c = 0$ at $\varphi > 10 \times 10^{19}$ e/cm$^2$ (see Fig. 2), while the resistivity increases and shows a metal-insulator transition above $\varphi_{10} \approx 11.2 \times 10^{19}$ e/cm$^2$. Accordingly, above about $\varphi_6 = 6.9 \times 10^{19}$ e/cm$^2$ the low-temperature resistivity starts to deviate upwards from the extrapolated linear dependence designated by the blue line in Fig. 1, indicating the onset of the metal-insulator transition and the disappearance of the PG. However, the slope $d\rho/dT$ in the normal state above $T^*$ remains virtually constant regardless of $\varphi$ (Table I). The constant slope $a = d\rho/dT$ of the resistivity curves in the normal state (Fig. 1) suggests that Matthiessen's rule is well satisfied in untwined YBCO single crystals [38,39].

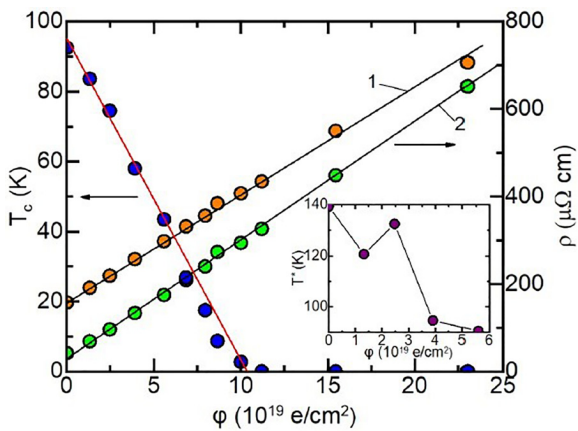

FIG. 2. In-plane resistivity $\rho$ (300 K) (1), $\rho$ (100 K) (2) and $T_c$ (3) of the optimally doped untwined YBa$_2$Cu$_3$O$_{6.96}$ single crystals at different irradiation doses from $\varphi_1 = 0$ to 1.3, 2.5, 3.9, 5.6, 6.9, 7.9, 8.7, 10.0, 11.2, 15.5 and $\varphi_{12} = 23 \times 10^{19}$ e/cm$^2$. Insert: the dose dependence of $T^*$ for the single crystal studied.

The evolution of the resistivity shown in Fig. 1, is well described by Matthiessen's formula, $\rho = \rho_0 + \rho_p$, where $\rho_0$ is the residual resistivity [4,33,49–51] and $\rho_p = aT$ is the temperature dependence of the resistivity of the pristine YBCO single crystal at $\varphi = 0$. Indeed, $\rho_0$ varies from $\rho_0 = -0.6\,\mu\Omega$ cm at $\varphi_1 = 0$ to $\rho_0 = 112.0\,\mu\Omega$ cm at $\varphi_{12} = 23 \times 10^{19}$ e/cm$^2$, whereas $d\rho/dT$ remains practically constant (Table I). The slope $d\rho/dT = 0.59 \pm 0.03\,\mu\Omega$/K (Table I) was calculated through a computer linear fitting of the experimental curves above $T^*$ and confirmed the linear behavior of $\rho(T)$ in the normal state with a root-mean-square error of $0.0018 \pm 0.0002$ over the specified temperature range for all $\varphi$.

Interestingly, Matthiessen's rule does not hold when the resistivity of YBCO varies greatly due to decreasing oxygen doping in single crystals [45] or in thin films [36] as well as due to irradiation defects in single crystals containing numerous twins and twin boundaries [31,35,37]. Thus, the observation of Matthiessen's rule turned out to be a characteristic feature of untwined YBCO single crystals and is an additional confirmation of the fact that twins are practically absent in our samples. The constant slop of $d\rho/dT$ also indicates the absence of the so-called S-shaped resistivity curves typical for the resistance behavior of underdoped YBCO with low $T_c$ and reduced $n_f$, the charge carrier density [45,49,52]. It is suggested [53] that the S-shaped form of the electrical resistance curves may arise due to the specific scattering of charge carriers at $T < T^*$ in HTSCs with the reduced $n_f$, possibly due to a partial modification of the Fermi surface [1,3,5,54]. Thus, the absence of S-shaped $\rho(T)$ curves for our single crystals with lower $T_c$ suggests that the oxygen content and $n_f$ in those crystals remain unchanged, despite the high defect content, present for high $\varphi$ [36,38]. The question arises: what is the physics behind the observed linear decrease in $T_c$ with increasing $\varphi$? As shown in Fig. 2, when $T_c$ decreases, both $\rho$ (300 K) and $\rho$ (100 K) increase with $\varphi$ also linearly. The resistivity is thought to increase due to the increase in disorder caused by irradiation defects [31,37–43]. Within the simple relativistic kinematic relation for electron-atom scattering it was found that the minimum electron energies required for displacement of O, Cu, Y, and Ba atoms are 129, 413, 532, and 730 keV, respectively, suggesting a threshold of 20 eV recoil energy [39]. Thus, there is no doubt that electrons with an energy of 2.5 MeV effectively displace all the atoms in the crystal, which leads to the appearance of many predominantly point or small cluster defects in the CuO$_2$ planes. This increases the disorder [31,36,38–43,48] and hence the scattering rate of charge carriers. However, assuming that the oxygen content and $n_f$ are independent on $\varphi$, as argued above, the linear decrease in $T_c$ still remains questionable [36,38].

To clarify the behavior of $T_c$, it is necessary to analyze the temperature dependences of excess conductivity at different $\varphi$. Since the excess conductivity and the PG disappear at approximately $\varphi_6 = 6.9 \times 10^{19}$ e/cm$^2$, as shown in Fig. 1, only five $\rho(T)$ curves with doses from $\varphi_1 = 0$ to $\varphi_5 = 5.6 \times 10^{19}$ e/cm$^2$ can be analyzed within the local pair model [4,5]. Figure 3 shows the temperature dependences of the resistivity $\rho(T) = \rho_{ab}(T)$ measured before irradiation $(\varphi_1 = 0)$ and for four subsequent irradiation doses, as specified in the figure caption. We are examining five different but similar in properties pristine single crystals, the properties of which are changed by irradiation (see Fig. 3 and Tables I, II, and III). As usual, $T_c$ of our samples is determined by extrapolating the linear part of $\rho(T)$ at the SC transition to its intersection with the $T$ axis [see Fig. 3(a)], which gives $T_c$ with an accuracy of $\sim 0.1$ K. As in [38], the width of the SC transition $\delta T_c$ is actually defined as the width of this linear region. As can be seen from Fig. 3, the $\delta T_c$ values measured in this way for all samples are quite narrow ($\delta T_c < 2.5$ K; see Table I), which indicates the good quality of our single crystal and the absence of additional phases with a different $T_c$. Despite the fact that $\delta T_c(\varphi)$ increases slightly under irradiation and has a maximum at $\varphi_3 = 2.5 \times 10^{19}$ e/cm$^2$ (Table I), the resistive transition at $\varphi_5 = 5.6 \times 10^{19}$ e/cm$^2$ still remains quite sharp, which indicates a uniform distribution of induced defects (disorder) over the sample volume [31,37,38]. Figure 3(b) shows, as an example, the exact definition of $T^*$ for $\varphi_3 = 2.5 \times 10^{19}$ e/cm$^2$ using the criterion $(\rho(T) - \rho_0)/aT = 1$, where $a = d\rho/dT$ [2]. This approach is used for all samples and allows one to determine the PG opening temperature $T^*$ with an accuracy of $\pm 0.3$ K. In contrast to the generally accepted phase diagram [3], the $T^*$ remains low and, moreover, decreases nonmonotonically with decreasing $T_c$, showing a pronounced anomaly (an increase of $T^*$) at $\varphi_3 = 2.5 \times 10^{19}$ e/cm$^2$ (see insert in Fig. 2). The possible reason for this feature will be discussed below. Thus, the observed temperature behavior of the resistivity (Figs. 1–3) is definitely different from what is usually observed with decreasing oxygen content and the density of charge carriers $n_f$ in YBCO [45,49]. Apparently, this unusual behavior seems to be typical for untwined YBCO single crystals [38,39].

TABLE I. Resistive and fluctuation conductivity parameters of $YBa_2Cu_3O_{7-\delta}$ single crystal at different irradiation doses. The parameters are described in the text.

| $\varphi$ ($10^{19}$ e/cm$^2$) | $\rho$ (300 K) ($\mu\Omega$ cm) | $\rho$ (100 K) ($\mu\Omega$ cm) | $d\rho/dT$ ($\mu\Omega$ cm/K) | $\rho_0$ ($\mu\Omega$ cm) | $\delta T_c$ (K) | $T_c$ (K) | $T_c^{mf}$ (K) | $T_G$ (K) | $T_0$ (K) | $T_{01}$ (K) |
|---|---|---|---|---|---|---|---|---|---|---|
| 0 | 157.8 | 41.8 | 0.56 | −6.0 | 0.7 | 92.5 | 92.90 | 93.0 | 93.2 | 96.0 |
| 1.3 | 187.2 | 68.9 | 0.59 | 15.0 | 1.0 | 83.6 | 84.45 | 84.6 | 85.0 | 90.0 |
| 2.5 | 218.8 | 96.2 | 0.59 | 41.6 | 4.3 | 74.5 | 75.37 | 75.7 | 79.0 | 88.7 |
| 3.9 | 257.0 | 133.5 | 0.62 | 72.0 | 2.3 | 58.0 | 59.12 | 59.9 | 63.5 | 71.2 |
| 5.6 | 297.0 | 175.0 | 0.62 | 112.0 | 2.2 | 43.5 | 44.77 | 45.1 | 46.1 | 51.0 |

The linear variation of $T_c$ with $\varphi$ (Fig. 2) was analyzed in detail in [38] by comparing the experimental results with the Abrikosov-Gorkov (AG) pair-breaking theory modified for d-wave superconductors [55], and the Emery-Kivelson (EK) theory, which proposes to take into account the phase fluctuations of the order parameter, which increase with decreasing so-called "phase stiffness" of the wave function [56]. The AG theory was shown to describes $T_c(\varphi)$ well at relatively low $\varphi$, but fails at high $\varphi$, most likely due to the very short coherence lengths $\xi(T)$ in HTSCs, which does not allow assuming a uniform disorder-averaged gap. On the other hand, the quantum phase fluctuations predicted by EK are found to be important for determining the actual $T_c$ in highly damaged material with low $T_c$, but are hardly applicable for superconductors with high $T_c$. So the main conclusion is that as the disorder increases, a crossover AG-EK from the pair-breaking regime to the phase fluctuation regime should occur [38]. However, surprisingly, no peculiarities are observed in the $\rho(\varphi)$ and $T_c(\varphi)$ dependencies (Fig. 2), and, strictly speaking, the reason for the linear decrease in $T_c$ in irradiated YBCO single crystals remained unclear. There is also no clear answer to the question of how the fluctuating Cooper pairs and PG behave under the influence of irradiation. Thus, to obtain new information, we studied the irradiation dose dependences of FLC and PG in the untwined YBCO single crystal, as described in following sections.

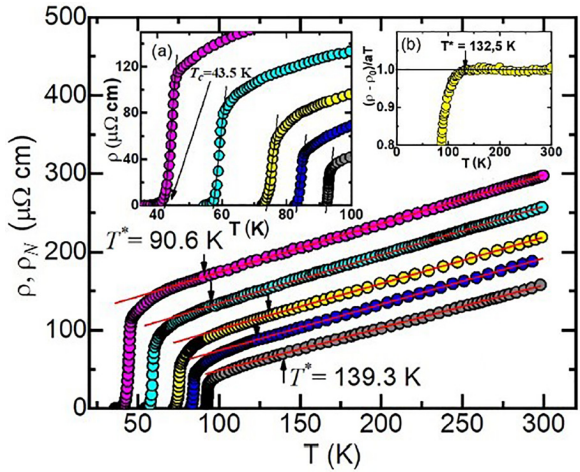

FIG. 3. Temperature dependences of the resistivity of an optimally doped $YBa_2Cu_3O_{6.96}$ single crystals ($T_c = 92.5$ K) at different irradiation doses $\varphi_1 = 0$ (gray dots), $\varphi_2 = 1.3 \times 10^{19}$ e/cm$^2$ (blue dots), $\varphi_3 = 2.5 \times 10^{19}$ e/cm$^2$ (yellow dots), $\varphi_4 = 3.9 \times 10^{19}$ e/cm$^2$ (turquoise dots), and $\varphi_5 = 5.6 \times 10^{19}$ e/cm$^2$ (magenta dots). Red lines denote $\rho_N(T)$ extrapolated linearly to lower temperatures. Arrows designate the PG resistive transitions to the superconducting state, which determine $T_c$ for single crystals at different $\varphi$. Inset (a) shows resistive transitions to the superconducting state, which determine $T_c$ for single crystals at different $\varphi$. Inset (b) shows the accurate determination of $T^*$ for $\varphi_3 = 2.5 \times 10^{19}$ e/cm$^2$ by using the criterion $(\rho(T) - \rho_0)/\alpha T = 1$, where $\alpha = d\rho/dT$ [2].

### B. Fluctuation conductivity

Excess conductivity, $\sigma'(T)$, in cuprate HTSCs is considered as the deviation of $\rho(T)$ below $T^* \gg T_c$ downward from its linear dependence at high temperatures and can be given by a simple formula:

$$\sigma'(T) = \sigma(T) - \sigma_N(T) = 1/\rho(T) - 1/\rho_N(T), \quad (1)$$

where $\rho_N(T) = aT + \rho_0$ is the resistivity of the sample in the normal state, extrapolated to the region of low temperatures. As before, $a = d\rho/dT$ determines the slope of the linear $\rho_N(T)$ dependence, and $\rho_0$ is the residual resistivity [49,57,58]. It should be borne in mind that at $T \lesssim T^*$ the DOS at the Fermi level begins to gradually decrease, indicating the opening of PG [5,7–9]. In addition, the Fermi surface is also thought to change [1,3–5,54] most likely due to the formation of local pairs (LPs) below $T^*$ [1,5,28], as mentioned above. It is well established that near $T_c$ LPs in cuprates behave like fluctuating Cooper pairs [5,49,59], but without long-range order (the so-called "short-range phase correlations") [12,18,28,59–63]. This is a rather specific HTSC state, characterized by a noticeable range of SC fluctuations $\Delta T_{\mathrm{fl}}$, which may occur due to the short coherence length $\xi(T)$ in combination with the quasi-two-dimensional (2D) structure of cuprates [5,12,28].

In accordance with the Hikami-Larkin theory [64] there is a three-dimensional (3D) state in HTSCs near $T_c$, where the coherence length along $c$ axis $\xi_c(T) \gg d$, and $d \approx 11.7$Å is the unit cell size in YBCO in the $c$-axis direction [57,63,65,66]. Many experiments have convincingly shown that in the 3D state, excess conductivity can be well described by the standard equation of the Aslamazov-Larkin theory for any 3D system (3D-AL) [5,57,63,67] (and references therein):

$$\sigma'_{3DAL} = C_{3D} \frac{e^2}{32\hbar\xi_c(0)} \varepsilon^{-1/2}, \quad (2)$$

where

$$\varepsilon = (T - T_c^{mf})/T_c^{mf}, \quad (3)$$

is the reduced temperature, $T_c^{mf}$ is the critical temperature in the mean-field approximation, as defined in Fig. 4 and will be discussed below, and $C_{3D}$ is the scaling factor. Interestingly, the better is the crystal structure of the sample, the closer $C_{3D}$ is to unity [4,49,65].

Since the HTSC coherence length above $T_c$ is given as $\xi(T) = \xi(0)\varepsilon^{-1/2}$ [63,68,69], it will decrease with increasing $T$. Ultimately, above the 3D-2D crossover temperature $T_0$ (see Fig. 4 and Table I) $d > \xi_c(T)$, and the 3D fluctuation state is lost. But in accordance with the theory [64,68] in a certain temperature range from $T_0$ to $T_{01}$ still $\xi_c(T) > d_{01}$, where $d_{01}$ ($\approx 3.6$ Å for YBCO [66]) is the distance between the CuO$_2$ conducting planes adjacent to the Y ions. Thus, up to the characteristic temperature $T_{01}$ (Table I), the planes form conducting blocks (double CuO$_2$ planes) coupled by Josephson interaction, forming a 2D fluctuation state [64,68]. In this case, in well-structured samples (thin films [63] and single crystals [69]), the excess conductivity is described by the 2D Maki-Thompson (2D-MT) theory, modified by Hikami and Larkin for HTSCs [64]:

$$\sigma'_{2DMT} = C_{2D} \frac{e^2}{8d\hbar} \frac{1}{1-\alpha/\delta} \ln\left(\frac{\delta}{\alpha} \frac{1+\alpha+\sqrt{1+2\alpha}}{1+\delta+\sqrt{1+2\delta}}\right) \varepsilon^{-1}, \quad (4)$$

where $\alpha = 2[\xi_c(0)/d]^2 \varepsilon^{-1}$ is the coupling parameter, and

$$\delta = \beta \frac{16}{\pi\hbar} \left[\frac{\xi_c(0)}{d}\right]^2 k_B T \tau_\phi \quad (5)$$

is the pair-breaking parameter. In accordance with the Hikami-Larkin theory at the 3D-2D crossover temperature $T_0 \sim \varepsilon_0$, $\delta \cong \alpha$, which allows us to obtain an equation for the lifetime of the fluctuating Cooper pairs $\tau_\phi \beta T = \pi\hbar/8k_B\varepsilon_0 = A/\varepsilon_0$, where $A = 2.998 \times 10^{-12}$ sK. The factor $\beta = 1.203(\ell/\xi_{ab})$, where $\ell$ is the mean-free path and $\xi_{ab}$ is the coherence length along CuO$_2$ planes, corresponds to the case of the clean limit ($\ell > \xi$) being typical for HTSCs [49,63,70].

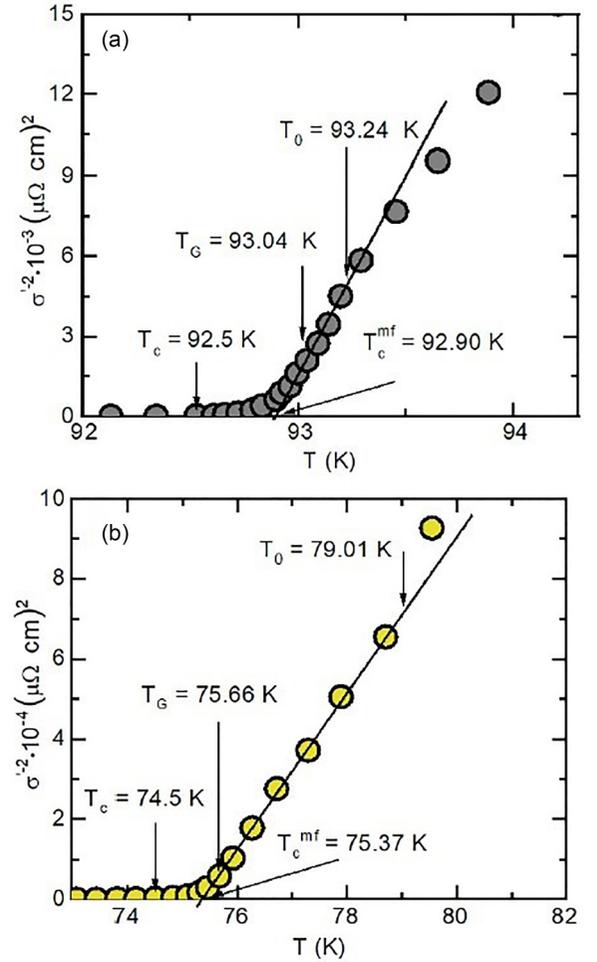

FIG. 4. $\sigma'^{-2}$ vs $T$ for the YBa$_2$Cu$_3$O$_{6.96}$ single crystals at two irradiation doses: (a) $\varphi_1 = 0$ e/cm$^2$ and (b) $\varphi_3 = 2.5 \times 10^{19}$ e/cm$^2$. Arrows point the characteristic temperatures: the mean-field transition temperature, $T_c^{mf}$, the superconducting transition temperature, $T_c$, the Ginzburg temperature, $T_G$, and 3D-2D crossover temperature, $T_0$. Straight lines correspond to the range of 3D-AL fluctuations. Different behavior of the FLC above $T_0$ is observed, caused by the influence of irradiation defects.

The 2D state is maintained until at $T_{01}$, above which $\xi_c(T) < d_{01}$, the data completely deviate from the theory [63,68]. This is because above $T_{01}$ any correlation between the superconducting layers is lost and the LPs are confined in the CuO$_2$ planes [68]. This results in a pronounced anisotropy of the coherence length in HTSCs, namely, $\xi_{ab}(0) \approx 14\xi_c(0)$ (Table II), where $\xi_c(0) = \xi_c(T^*) \approx 1$ Å [57,63,65], which is very small, even smaller than the characteristic dimension of the crystal structure $d_{01}$ along the $c$ axis. Thus, $T_{01}$ is the temperature that limits from above the range of SC fluctuation $\Delta T_{\mathrm{fl}} = T_{01} - T_G$ [4,5,64,68]. Accordingly, $T_G$ is the Ginzburg temperature, down to which the mean-field theory operates [71–73]. So it turns out that fluctuation conductivity is only part of complete $\sigma'(T)$ given by Eq. (1) and is realized in a relatively narrow range of SC fluctuations from $T_G$ to $T_{01}$, e.g., $\Delta T_{fl} = 3$ and 13 K, for not irradiated and irradiated with $\varphi_3 = 2.5 \times 10^{19}$ e/cm$^2$ single crystals, respectively (see Table II). A graphical illustration of the definitions of all the characteristic temperatures we are talking about is presented in Fig. 4. Obviously, at $T = T_0$, $\xi_c(T_0) = \xi_c(0)\varepsilon_0^{-1/2} = d$ and $\xi_c(0)$ can be easily determined by the simple formula [5,49,63]

$$\xi_c(0) = d\sqrt{\varepsilon_0}.\qquad(6)$$

TABLE II. Fluctuation conductivity parameters of YBa$_2$Cu$_3$O$_{7-\delta}$ single crystal at different irradiation doses. The parameters are described in the text.

| $\varphi$ ($10^{19}$ e/cm$^2$) | $\Delta T_{fl}$ (K) | $C_{3D}$ | $d_{01}$ (Å) | $\xi_c(0)$ (Å) | $\xi_{ab}(0)$ (Å) |
|---|---|---|---|---|---|
| 0 | 3.0 | 0.79 | 3.9 | 0.71 ± 0.02 | 10± 0.02 |
| 1.3 | 5.4 | 0.98 | 4.1 | 1.05 ± 0.02 | 14.8± 0.02 |
| 2.5 | 13.0 | 2.82 | 6.1 | 2.57 ± 0.02 | 36.2± 0.02 |
| 3.9 | 11.3 | 2.25 | 7.1 | 3.19 ± 0.02 | 44.9± 0.02 |
| 5.6 | 5.9 | 0.77 | 5.3 | 1.98 ± 0.02 | 27.9± 0.02 |

In turn, at $T = T_{01}$, $\xi_c(T_{01}) = \xi_c(0)\varepsilon_{01}^{-1/2} = d_{01}$. Then, using Eq. (6), we calculated $d_{01} = d\sqrt{\varepsilon_0}/\sqrt{\varepsilon_{01}}$ for all single crystals studied, and the results are listed in Table II.

Interestingly, in compounds with imperfect structure (ceramics [74], first thin films [65], and films with artificially produced defects [75]) defects eliminate all manifestations of FLC we have described. As a result, in the case of structurally disturbed structure, the excess conductivity is described by the Lawrence-Doniach (LD) model [76], developed for layered superconductors:

$$\sigma'_{LD} = C_{LD}\frac{e^2}{16\hbar d\sqrt{1+2\alpha}}\varepsilon^{-1},\qquad(7)$$

where $\alpha$ is the coupling parameter, as noted above. In this case $\sigma'$ will diverge as $\varepsilon^{-1/2}$ (3D behavior) when the temperature is close to $T_c^{mf}$, and $\sigma'$ will change as $\varepsilon^{-1}$ (2D behavior) at sufficiently higher temperatures such that $\alpha = 2[\xi_c(0)/d]^2\varepsilon^{-1} < 1$. Thus, Eq. (7) gives a long smooth transition from 3D to 2D behavior without any pronounced features [65,75], which does not allow one to determine any parameters of the sample structure, in contrast to the case analyzed in this work. It is clear lead to irradiation with high-energy electrons of 2.5 MeV will lead to the formation of many defects in the initial crystal (see Figs. 1 and 3), which should somehow change the FLC properties. To assess the influence of defects in the case under consideration, it is necessary to trace which of the listed theories describes the FLC and how this description changes with increasing $\varphi$.

Outside the critical region, $\sigma'(T)$ is a function only of the reduced temperature $\varepsilon$ [Eq. (3)], where $T_c^{mf}$ is the mean-field transition temperature; hence determination of $T_c^{mf}$ is of primary importance for determining both $\sigma'(T)$ and PG values $\Delta^*(T)$ [4,5,57,63,68,69]. $T_c^{mf}$ is the critical temperature in the mean-field approximation, which limits from above the range of critical fluctuations around $T_c$, where the SC order parameter $\Delta < k_B T$ [71–73]. $T_c^{mf}$ is determined by extrapolating to zero the linear 3D region of the $\sigma'^{-2}$ vs $T$ plot (Fig. 4), since as $T$ approaches $T_c^{mf}$ from higher temperatures, $\sigma'$ must diverge as $(T-T_c^{mf})^{-1/2}$ [65]. Note that $T_c^{mf} > T_c$, and typically the shift is of 1–2 K, which is a rough measure of the size of the critical region above $T_c$ [65,72,73]. The corresponding plots $\sigma'^{-2}$ vs $T$ for $\varphi_1 = 0$ and $\varphi_3 = 2.5 \times 10^{19}$ e/cm$^2$, as an example, are shown in Fig. 4. In addition to $T_c^{mf}$, all other characteristic temperatures, $T_c$, $T_G$, and the 3D-2D crossover temperature $T_0$, are also shown. Using this approach, $T_c^{mf}$ and other specific temperatures were obtained for all five samples and are listed in Table I. In Fig. 4, the linear part of $\sigma'^{-2}(T)$, corresponding to the 3D FLC, is clearly visible. But above $T_0$, for $\varphi_1$, the data deviate to lower values indicating the presence of 2D-MT fluctuations [63,68], whereas for $\varphi_3$, the data deviate to higher values, indicating complete suppression of 2D-MT fluctuations by disorder [76]. Interestingly, at $\varphi_5 = 5.6 \times 10^{19}$ e/cm$^2$ with $T_c^{mf} = 44.77$ K the data again deviate to the right from the $\sigma'^{-2}(T)$ straight line, suggesting a recovery of 2D-MT fluctuations despite the huge number of defects.

To clarify this unusual behavior, we traditionally plot $\ln\sigma'$ as a function of $\ln\varepsilon$ for each single crystal irradiated with electrons at doses from $\varphi_1$ to $\varphi_5$ and compare the plots with theories, as shown in Fig. 5. Note that with increasing $\varphi$ (increasing number of irradiation defects), the absolute value of $\sigma'(T)$ decreases significantly. Nevertheless, in the 3D state, near $T_c$, the experimental data are well described by the 3D-AL theory [Eq. (2)], regardless of $\varphi$. In the double logarithmic scale, the predictions of the theory are shown as the red solid lines with slope $\lambda = -1/2$, denoted in the figure as 3D-AL. At temperatures above $T_0$ ($\ln\varepsilon_0$ in the figures), which limits the range of 3D fluctuations from above, a clear crossover to the 2D fluctuations regime is observed for all $\varphi$. In the case of $\varphi_1 = 0$ (gray dots 1) and $\varphi_2 = 1.3 \times 10^{19}$ e/cm$^2$ (dark blue dots 2), $\sigma'(\varepsilon)$ above $T_0$ is well described by the 2D-MT fluctuation contribution calculated using [Eq. (4)], which is typical for well-structured HTSCs [63,69]. Interestingly, at $\varphi_2$, despite the fact that the total number of defects is believed to increase significantly, the contribution of 2D-MT fluctuations is even more pronounced than at $\varphi_1 = 0$. In addition, in this case the scaling factor $C_{3D} \approx 1$ (Table II), as in the case of a well-structured YBCO films [4,49,63]. Strictly speaking, both findings seem somewhat unexpected. Moreover, as can be seen in the figure, the LD model (green dashed curve) is in complete disagreement with the experimental results, suggesting, somewhat surprisingly, that the sample still has a relatively good structure even under irradiation. Finally, above $T_{01}$, denoted as $\ln\varepsilon_{01}$ in Fig. 5, the experimental data completely deviate downwards from any of the theories considered.

Knowing all the characteristic temperatures, $T_c^{mf}$, $T_G$, and $T_0$ from $\sigma'^{-2}(T)$ data, as shown in Fig. 4, and $T_{01}$ from $\ln\sigma'(\ln\varepsilon)$ data, as presented in Fig. 5, we were able to determine such sample parameters as $\Delta T_{fl}$, $\xi_c(0)$, $d_{01}$, and $C_{3D}$ for all single crystals studied. As can be seen from Table II, all these parameters increase significantly at $\varphi > 2.0 \times 10^{19}$ e/cm$^2$, indicating the significant increase of disorder, which in turn leads to noticeable changes in the behavior of the FLC, as shown in Fig. 5(b). Indeed, the data convincingly show that at $\varphi_3 = 2.5 \times 10^{19}$ e/cm² (yellow dots 3) and $\varphi_4 = 3.9 \times 10^{19}$ e/cm² (turquoise dots 4) the 2D-MT fluctuations are completely suppressed by the irradiation defects. Now, somewhat unexpectedly, the 2D fluctuations are well described by the well-known AL equation for 2D superconductors (the 2D-AL equation), which describes well their FLC near $T_c$ [67], including HTSCs [68,69]:

$$\sigma'_{2DAL} = C_{2D}\frac{e^2}{16\hbar d}\varepsilon^{-1}, \qquad (8)$$

as shown in Fig 5(b) by the purple solid lines with slope $\lambda = -1$ and labeled 2D-AL. Surprisingly, the typical for large disorder LD model (Eq. (7)) [75,76] (dashed green curve) still does not match the experimental results. In Fig. 5, the LD curves are shown only for $\varphi_2$ and $\varphi_5$ to avoid clutter, but the LD curves for $\varphi_1$, $\varphi_3$, and $\varphi_4$ also turned out to be incompatible with the experiment. In turn, Eq. (8) describes the experimental results fairly well in a wide temperature range from $T_0$ to $T_{01}$ ($\ln\varepsilon_0$ and $\ln\varepsilon_{01}$ in the figure), taking $d = 11.7$ Å, the unit cell size for YBCO in the $c$ direction [66]. This leads to a noticeable rise in the range of SC fluctuations $\Delta T_{fl}$, which increases by more than four times for $\varphi_3 = 2.5 \times 10^{19}$ e/cm² (Table II). To our knowledge, 2D-AL fluctuations, especially over such a wide temperature range, have not been observed in cuprates before. Accordingly $d_{01}$ also increases significantly (see Table II), most likely due to a decrease in the thickness of the Ba-O layers in the YBCO crystal as a result of the displacement of O and Cu atoms during irradiation, leading to a decrease in the distance between Cu1 and Cu2 [66]. Simultaneously, the coherence length $\xi_c(T) = \xi_c(0)\varepsilon^{-1/2}$ also increases, due to the more than fourfold increase of $\xi_c(0)$ at $\varphi_4$ (see Table II). However, the necessary two-dimensionality condition $d > \xi_c(T) > d_{01}$ [64,68] is still satisfied in the entire temperature range from $T_0$ to $T_{01}$ for all $\varphi$.

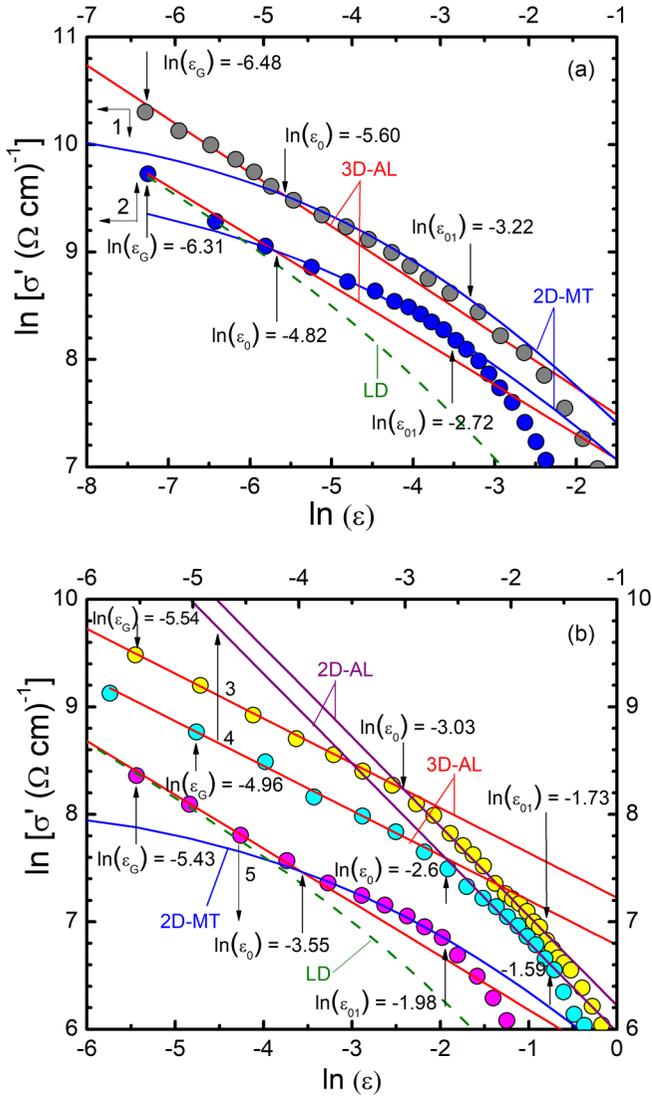

FIG. 5. $\ln\sigma'$ vs $\ln\varepsilon$ for $YBa_2Cu_3O_{6.96}$ single crystals: (a) at $\varphi_1 = 0$ (curve 1, gray dots) and $\varphi_2 = 1.3 \times 10^{19}$ e/cm² (curve 2, blue dots) and (b) at $\varphi_3 = 2.5 \times 10^{19}$ e/cm² (curve 3, yellow dots), $\varphi_4 = 3.9 \times 10^{19}$ e/cm² (curve 4, turquois dots), and $\varphi_5 = 5.6 \times 10^{19}$ e/cm² (curve 5, magenta dots) in comparison with fluctuation theories: 3D-AL (red lines), 2D-MT (blue curves), 2D-AL (purple lines), and the LD model (dashed green curve). All characteristic temperatures are marked with arrows: $\ln(\varepsilon_G)$ refers to the Ginzburg temperature $T_G$, $\ln(\varepsilon_0)$ to the 3D-2D crossover temperature $T_0$, and $\ln(\varepsilon_{01})$ to $T_{01}$, which limits the region of the SC fluctuations from above. The $x$-axis scales are shifted to avoid data overlap. In panel (b) the upper $x$ scale is used for $\varphi_3$ and $\varphi_4$, while the lower $x$ scale is used for $\varphi_5$.

Naturally, one would expect that with an increase in the irradiation dose to $\varphi_5 = 5.6 \times 10^{19}$ e/cm², the contribution of 2D-AL fluctuations would become even more pronounced. However, in this case, a completely unexpected result is obtained [Fig. 5(b), magenta dots], namely, above $T_0$ the experimental points deviate upwards from the extrapolated 3D-AL dependence (red line) and are again well described by the 2D-MT theory (blue curve). Surprisingly, it turns out that by creating many defects, we eventually obtained the $\ln\sigma'(\ln\varepsilon)$ dependence, typical for HTSC without defects, approximately the same as in Fig. 5(a). The first thing that comes to mind is that the crystal structure of the sample has somehow been restored. The more defects, the more isotropic the sample [31]. This is also evidenced by the sharp decrease in all parameters of the FLC analysis: $\Delta T_{fl}$, $\xi_c(0)$, $d_{01}$, and especially $C_{3D} = 0.77$, which have become almost the same as in the case of $\varphi_1$ and $\varphi_2$ (see Table II). It can therefore be concluded that the effect of strong electron irradiation on fluctuating Cooper pairs in HTSC is very specific and differs greatly from the effect of irradiation on transport properties (Figs. 1 –3). However, it should be borne in mind that the fluctuation theories of AL and MT describe the FLC in a relatively narrow temperature range: approximately 15 K above $T_c$. To obtain information on the behavior of excess conductivity under irradiation in the entire temperature range from $T_G$ to $T^*$, it is necessary to consider the evolution of the temperature dependences of PG at different irradiation doses.

## C. Temperature dependence of the pseudogap

As mentioned in the Introduction, the pseudogap (PG) is one of the most attractive and mysterious properties of HTSCs. As stated, we believe that the PG arises below $T^*$ due to the formation of paired fermions, the so-called local pairs (LPs), which are transformed into fluctuating Cooper pairs near $T_c$ [5,21–25,59–63]. It is believed that if the processes leading to the opening of the PG at $T^*$ did not occur in HTSC, then $\rho_N(T)$ would remain linear down to $\sim T_c$ [1,5,63,69]. Thus, it is obvious that the excess conductivity $\sigma'(T)$ arises as a result of the opening of PG and therefore should contain information about its magnitude and temperature dependence, which has to change under irradiation [1,5,43,49,57]. To obtain such information, we need an equation that defines the entire experimental curve $\sigma'(T)$ from $T^*$ to $T_G$ and contains the PG parameter $\Delta^*(T)$ explicitly. In addition, the dynamics of LPs formation, which is proportional to $(1 - T/T^*)$, and the pair-breaking effect, described by the factor [exp$(-\Delta^*/T)$], should also be taken into account to correctly describe our experimental results. Due to the lack of a rigorous fundamental theory, the equation for $\sigma'(T)$ was proposed in [49]:

$$\sigma'(T) = A_4 \frac{e^2\left(1 - \frac{T}{T^*}\right)\exp\left(-\frac{\Delta^*}{T}\right)}{16\hbar\xi_c(0)\sqrt{2\varepsilon^*_{c0}}\sinh(2\varepsilon/\varepsilon^*_{c0})}, \quad (9)$$

where $A_4$ is a scaling factor, the same as the $C$ factor in the FLC theory [see Eqs. (2), (4), (7), and (8)]. The temperature of the pseudogap opening $T^*$, the reduced temperature $\varepsilon$ [Eq. (3)] and the coherence length $\xi_c(0)$ [Eq. (6)] have already been determined from the resistivity and the FLC analysis (Tables I and II). Other necessary parameters, such as the pseudogap $\Delta^*(T_G)$ [63,77–79], the parameter $\varepsilon*_{c0}$ [80], and the coefficient $A_4$ can now be found directly from our experiment within the framework of the analysis developed in [49].

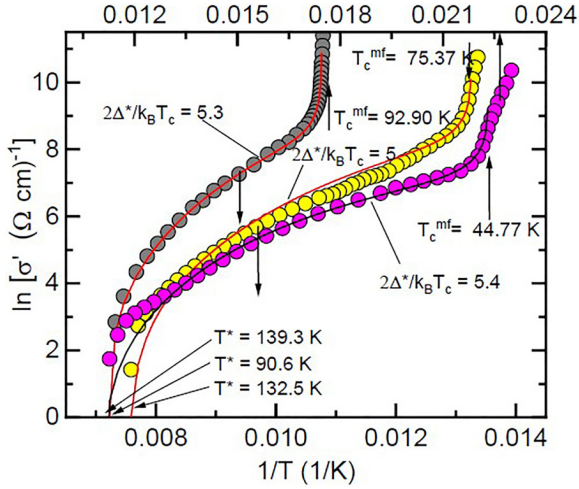

FIG. 6. ln$\sigma'$ vs $1/T$ for YBa$_2$Cu$_3$O$_{6.96}$ single crystals at $\varphi_1 = 0$ (gray dots), $\varphi_3 = 2.5 \times 10^{19}$ e/cm$^2$ (yellow dots), and $\varphi_5 = 5.6 \times 10^{19}$ e/cm$^2$ (magenta dots) plotted in the temperature range from $T^*$ down to $T_c^{mf}$. The solid curves are fits to the data with Eq. (9). The best fits are obtained for $\Delta^*(T_G)$ resulting in $D^* = 2\Delta^*(T_G)/k_B T_c =$ 5.3, 5.0, and 5.4 for $\varphi_1$, $\varphi_3$, and $\varphi_5$, respectively. The $x$ axis is shifted for readability. The upper $x$ axis is used for $\varphi_5$.

To find $\Delta^*(T_G)$, we plot the graphs of the experimental excess conductivity ln$\sigma'$ vs $1/T$ and approximate them by the ln$\sigma'(1/T)$ dependence calculated using Eq. (9) [5,63,69]. Figure 6 shows such plots for single crystals irradiated with $\varphi_1 = 0$, $\varphi_3 = 2.5 \times 10^{19}$ e/cm$^2$ and $\varphi_5 = 5.6 \times 10^{19}$ e/cm$^2$. Since the difference in $T_c$ with $\varphi_1$ is very large, a different upper $x$ axis is used for $\varphi_5$. When plotted in these coordinates, the shape of the theoretical curves turns out to be very sensitive to the value of $\Delta^*(T_G)$ (see, for example, Fig. 9 in [63]). It should also be borne in mind that in cuprates $\Delta^*(T_G) = \Delta(0)$, which is the SC energy gap at $T = 0$ [78,79]. Taking this into account, we can determine the BCS ratio $D^* = 2\Delta(0)/k_B T_c = 2\Delta^*(T_G)/k_B T_c$ and, consequently, calculate $D^*$ values using $\Delta^*(T_G)$, which results in the best fitting to the experimental data (see Fig. 6). So, for example, $D^* = 5.3$ for $\varphi_1$, $D^* = 5.0$ for $\varphi_3$ and $D^* = 5.4$ for $\varphi_5$ (Table III). $D^* = 5 \pm 0.2$ is a typical value for YBCO [81], which suggests a strong-coupling limit for our samples [70], despite the strong influence of disorder on $T_c$. The shape of the ln$\sigma'(1/T)$ curves for $\varphi_2 = 1.3 \times 10^{19}$ e/cm$^2$ and $\varphi_4 = 3.9 \times 10^{19}$ e/cm$^2$ looks exactly the same, but the curves are not shown to avoid cluttering the graph.

It has been established that in all cuprates $\sigma'^{-1} \sim \exp(\varepsilon)$ in a certain temperature range ln$\varepsilon_{c01} < $ ln$\varepsilon < $ ln$\varepsilon_{c02}$ above $T_{01}$ (Fig. 7) [80]. On the main panel the corresponding ln$\varepsilon_{c01}$ and ln$\varepsilon_{c01}$ for different samples are marked with arrows of different length. Accordingly, for the not irradiated single crystal ($\varphi_1 = 0$), in the temperature range $\varepsilon_{c01} < \varepsilon < \varepsilon_{c02}$ (103.8 K $< T < $ 116.1 K), ln($\sigma'^{-1}$) is a linear function of $\varepsilon$ with a slope $\alpha^* = 8.5$ (inset, Fig. 7), which determines the parameter $\varepsilon^*_{c0} = 1/\alpha^* \approx 0.12$. This approach allows us to determine reliable values of $\varepsilon*_{c0}$ for all samples (Table III), which at high temperatures significantly affect the curvature of the theoretical $\Delta^*(T)$ curves presented in Fig. 8 and $\sigma'(T)$ curves shown in Figs. 6 and 7. It is worth emphasizing that the dependence $\sigma'^{-1} \sim \exp(\varepsilon)$ is observed in the temperature range around $T_{pair}$, where SBBs are transformed into local pairs.

To find the scaling factor $A_4$, we plot ln$\sigma'$ vs ln$\varepsilon$ for all samples in the entire temperature range from $T^*$ to $T_G$ (Fig. 7, main panel) and approximate them with Eq. (9) with the parameters already found (see the tables). We fit the theoretical curves to the experimental data in the region of 3D-AL fluctuations, where ln$\sigma'$ is a linear function of ln$\varepsilon$ with slope $\lambda = -1/2$ (Fig. 7, red and black curves) [49]. Note that now the only fitting parameter is $A_4$. Very good agreement between the experimental data and our theoretical model appears for all $\varphi$ using the $A_4$ values listed in Table III, as shown in Fig. 7 for single crystals irradiated with doses of $\varphi_1$, $\varphi_3$, and $\varphi_5$. In the case of $\varphi_3$, the theoretical curve in the range of 2D fluctuations lies slightly above the data, since the 2D-MT fluctuations are completely suppressed, as noted above. The curves for $\varphi_2 = 1.3 \times 10^{19}$ e/cm$^2$ and $\varphi_4 = 3.9 \times 10^{19}$ e/cm$^2$ are not shown in the figure to avoid cluttering the graph. As expected, $A_4$ increases markedly with increasing disorder, showing the highest value of 82 at $\varphi_4$, but then surprisingly sharply decreases, showing a minimum value of 20.5 at $\varphi_5$ (see Table III). Accordingly, a very regular shape of the experimental curve is observed at $\varphi_5$ (Fig. 7), with a pronounced fluctuation contribution from 2D-MT as well as the extremely good agreement with the theory [see also the corresponding data in Fig. 5(b)]. This result looks quite surprising, as already mentioned above, since in this case the number of irradiation defects should be very large and strong damage to the sample structure should be expected. To shed more light on this unexpected result, the properties of PG under irradiation will now be examined in detail.

TABLE III. Pseudogap parameters of YBa$_2$Cu$_3$O$_{7-\delta}$ single crystals at different irradiation doses. The parameters are described in the text.

| $\varphi$ ($10^{19}$ e/cm$^2$) | $T^*$ (K) | $T_{pair}$ (K) | $\varepsilon^*_{c0}$ | $A_4$ | $D^*$ | $\Delta^*(T_G)/k_B$ (K) | $\Delta^*(T_{pair})/k_B$ (K) |
|---|---|---|---|---|---|---|---|
| 0 | 139.3 | 114.0 | 0.12 | 34 | 5.3 | 245.8 | 246.7 |
| 1.3 | 120.6 | 99.8 | 0.14 | 39 | 5.0 | 209.3 | 199.5 |
| 2.5 | 132.5 | 91.3 | 0.15 | 74 | 5.0 | 184.2 | 220.5 |
| 3.9 | 94.7 | 73.6 | 0.14 | 82 | 5.4 | 152.5 | 160.8 |
| 5.6 | 90.6 | 68.6 | 0.20 | 20.5 | 5.4 | 117.5 | 118.4 |

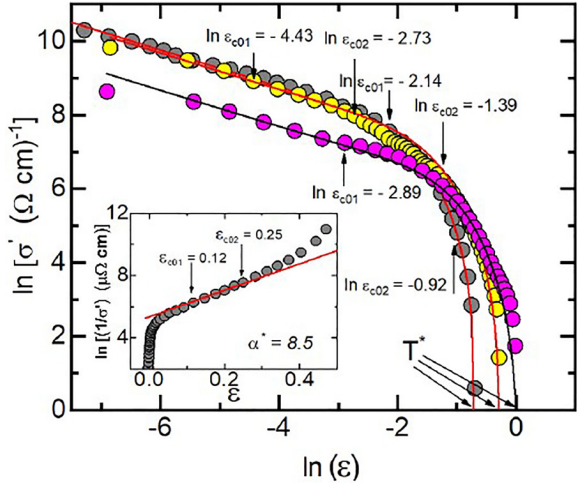

FIG. 7. $\ln \sigma'$ vs $\ln \varepsilon$ for YBa$_2$Cu$_3$O$_{6.96}$ single crystals at $\varphi_1 = 0$ (gray dots), $\varphi_3 = 2.5 \times 10^{19}$ e/cm$^2$ (yellow dots), and $\varphi_5 = 5.6 \times 10^{19}$ e/cm$^2$ (magenta dots), plotted over the temperature range from $T^*$ down to $T_G$ and compared with our theoretical model [Eq. (9)] (red and black curves). The arrows designate the ranges of the dependences where $\sigma'^{-1} \sim \exp(\varepsilon)$. Inset: $\ln \sigma'^{-1}$ as a function of $\varepsilon$ for $\varphi_1 = 0$. Solid line indicates the linear part of the curve between $\varepsilon_{c01} = 0.12$ and $\varepsilon_{c02} = 0.25$. Its slope $\alpha^* = 8.5$ determines the parameter $\varepsilon *_{c0} = 1/\alpha^* \approx 0.12$. The $\varepsilon *_{c0}$ and other relevant parameters are given in Table III for single crystals at all irradiation doses.

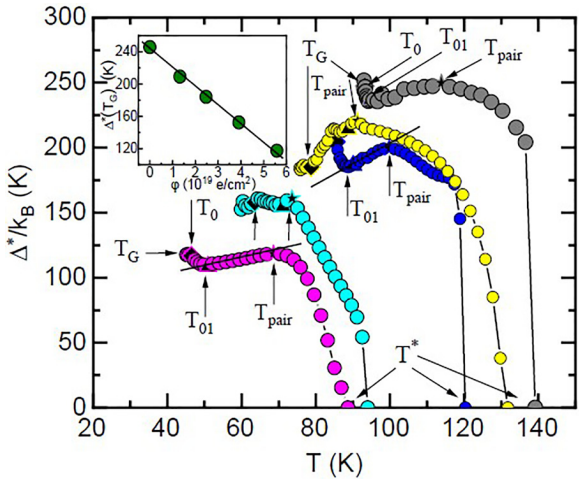

FIG. 8. Temperature dependencies of pseudogap $\Delta^*$ for YBa$_2$Cu$_3$O$_{6.96}$ single crystals at $\varphi_1 = 0$ (gray dots), $\varphi_2 = 1.3 \times 10^{19}$ e/cm$^2$ (blue dots), $\varphi_3 = 2.5 \times 10^{19}$ e/cm$^2$ (yellow dots), $\varphi_4 = 3.9 \times 10^{19}$ e/cm$^2$ (turquois dots), and $\varphi_5 = 5.6 \times 10^{19}$ e/cm$^2$ (magenta dots) calculated using Eq. (10) with parameters as specified in the text. Arrows indicate the corresponding characteristic temperatures. Stars point the maxima close to $T_{pair}$. Inset shows $\Delta^*(T_G)$ vs $\varphi$ in the range from $\varphi_1$ to $\varphi_5$. Solid curves are guidance for eye.

Solving Eq. (9) for the pseudogap, $\Delta^*(T)$, the following formula is obtained [49]:

$$\Delta^*(T) = T\ln\left[A_4\left(1 - \frac{T}{T^*}\right)\frac{1}{\sigma'(\varepsilon)}\frac{e^2}{16\hbar\xi_c(0)}\right.$$
$$\left. \times \frac{1}{\sqrt{2\varepsilon *_{c0}}\sinh(2\varepsilon/\varepsilon *_{c0})}\right], \quad (10)$$

where $\sigma'(\varepsilon)$ is the measured excess conductivity. The excellent approximation of the data by Eq. (9) (Fig. 7) gives reason to believe that Eq. (10) with the sets of parameters we found (Tables I–III) will give the correct values and temperature dependences of the PG, $\Delta^*(T)$. Figure 8 shows $\Delta^*(T)$ for single crystals irradiated with doses $\varphi_1 - \varphi_5$, as calculated using Eq. (10). For example, $\Delta^*(T)$ for $\varphi_1 = 0$ is calculated with the following set of parameters: $T^* = 139.3$ K, $T_c^{mf} = 92.9$ K, $\xi_c(0) = 0.71$ Å, $\varepsilon *_{c0} = 0.12$, and $A_4 = 34$ and is represented by gray dots in the figure. As expected, in this case the shape of the $\Delta^*(T)$ dependence with the highest $T^* = 139$ K, broad maximum at $T_{pair} = 114$ K, which corresponds to the BEC-BCS crossover [28,49], pronounced minimum at $T_{01} = 96.0$ K, maximum around $T_0 \approx 93.2$ K, and a small minimum at $T_G = 93.0$ K, is typical for optimally doped YBCO single crystals without twins [69].

The influence of electron irradiation on the properties of the single crystals becomes clearly noticeable already for the dose $\varphi_2 = 1.3 \times 10^{19}$ e/cm$^2$. Indeed, $\rho$ (100 K) increased by $\sim$1.6 times, and all characteristic temperatures (Table I), $\sigma'(T)$ [Fig. 5(a)] and especially $\Delta^*(T_G)$ and $\Delta^*(T_{pair})$ (Table III) decreased significantly, while $\Delta T_{fl}$, $d_{01}$ and $\xi_c(0)$ changed only slightly (Table II). As a result, the shape of $\Delta^*(T)$ still looks quite uniform, and the features at all characteristic temperatures $T_{pair}$, $T_0$ and $T_G$ are very pronounced, suggesting that in this case the effect of defects on FLC [Fig. 5(a)] and PG (Fig. 8) is relatively small [82]. This is confirmed by the observation of a pronounced 2D-MT fluctuation contribution [Fig. 5(a)] with $C_{3D} \approx 1.0$ (Table II), being typical for well-structured YBCO, as mentioned above. Thus, at relatively low irradiation doses (such as $\varphi_2$), $\Delta^*(T)$ behaves as expected.

The main specific feature in the behavior of $\Delta^*(T)$ is observed for the single crystal irradiated with $\varphi_3$ (see Fig. 8), which is approximately twice as large as $\varphi_2$. In this case, instead of the expected decrease, a rather unexpected increase in $\Delta^*(T)$ is observed, accompanied by an increase in $T^*$ by $\sim 12$ K and $\Delta^*(T_{\text{pair}})/k_B$ by $\sim 20$ K (Table III). Accordingly, the $\Delta^*(T)$ curve became $\sim 1.2$ times wider than even at $\varphi_1$ and shows the largest width $(T^* - T_G) = 56.8$ K. As a result, the experimental data $\Delta^*(T)$ lay above those obtained for the single crystal at $\varphi_2$. Taking into account these results and analyzing the data presented in Fig. 2 in [38], we come to the conclusion that $\varphi_3$ corresponds to the irradiation dose at which the AG-EK crossover occurs, that is, the mechanism of interaction of irradiation with local pairs changes. This assumption allows us to justify the observed increase in $\Delta^*(T)$ at $\varphi_3$. The corresponding analysis shows that in the whole temperature range (85 – 110 K), where $\Delta^*(T)$ at $\varphi_3$ exceeds $\Delta^*(T)$ at $\varphi_2$ (Fig. 8), $\sigma'(T)$ at $\varphi_3$ is noticeably smaller than at $\varphi_2$. As Fig. 5(b) shows, such a decrease in $\sigma'(T)$ occurs after the change of the $\sigma'(T)$ dependence from 2D-MT to 2D-AL above $T_0$ at $\varphi_3$. Since $\Delta^*(T) \sim \ln[1/\sigma'(T)]$ [Eq. (10)], this should ensure a sharp increase in $\Delta^*(T)$ observed at $\varphi_3$.

With a further increase in the irradiation dose to $\varphi_4 = 3.9 \times 10^{19}$ e/cm$^2$, the strong decrease in $\Delta^*(T_{\text{pair}})/k_B$ by $\sim 60$ K compared to obtained for the single crystal with $\varphi_3$ is observed (Table III), which is consistent with the expected enhancement of EK phase fluctuations by irradiation defects [38]. As a result, a strong decrease in $T^*$ (see Fig. 3 and Table III) occurs, and hence the temperature range in which the calculations are carried out. Thus, in this case we observe a very strong effect of irradiation, leading to high precision with the largest $\xi_c(0) \approx 3.2$ Å and $d_{01} \approx 7.1$ Å (Table II), which is about twice as large as reported for well-structured YBCO [66]. At the same time, the width $(T^* - T_G)$ also decreases significantly and demonstrates the smallest value of $\sim 34.8$ K. Eventually, $\Delta^*(T)$ takes on a very unusual narrow shape, which we have never observed before.

It was natural to expect that with an increase in the irradiation dose to $\varphi_5 = 5.6 \times 10^{19}$ e/cm$^2$, which is more than 1.4 times larger than $\varphi_4$, leading to a further increase in disorder, the shape of $\Delta^*(T)$ will be similar to that obtained at $\varphi_4$, only all dimensions of $\Delta^*(T)$ will be smaller. However, as can be seen from Fig. 8, the results we obtained were completely different. The maximum value of $\Delta^*(T_{\text{pair}})/k_B$ has really decreased by $\sim 40$ K, but the width $(T^* - T_G)$ of $\Delta^*(T)$ has clearly increased and is equal to 45.5 K (Tables I and III). Moreover, a linear part between $T_{\text{pair}} = 68.6$ K and $T_{01} = 51.0$ K is observed, which slope $\alpha = 0.55$ is almost identical ($\alpha = 0.53$) to that we observed in optimally doped YBCO single crystals without any irradiation defects [82]. Thus, it can be concluded that, according to the results of the FLC analysis [Fig. 5(b)], a very uniform and smooth $\Delta^*(T)$ curve is observed with clear maximum at $T_{\text{pair}}$, pronounced minimum at $T_{01}$, maximum close to $T_0$, and small minimum at $T_G$ (Fig. 8). Such a $\Delta^*(T)$ dependence is typical for optimally doped YBCO single crystals with a small number of defects, what is very surprising.

At first glance, this result is completely unexpected and cannot be explained within the standard resistive approach, especially taking into account the huge number of defects produced by irradiation. It is obvious that irradiation-induced defects greatly increase the scattering of normal charge carri-ers. As a result, the crystal resistance increases sharply, and at $\varphi_6$, as already noted, the metal-insulator transition begins (Fig. 1). However, the number of local pairs in the sample also apparently remains quite large. As can be seen from the experiment, superconductivity is preserved up to very high irradiation doses, practically up to $\varphi_{10}$ (Figs. 1 and 2). Thus, we are still dealing with two subsystems of charge carriers in HTSC, namely, normal charge carriers and local pairs, most likely in the form of fluctuating Cooper pairs (FCPs) near $T_c$. However, the influence of defects on the FCPs, which by definition transfer charge without dissipation, is not so obvious.

Apparently, it should be borne in mind that, as already noted, the more defects, the more isotropic the sample [31,39]. It is quite probable that at $\varphi_5$ the number of induced defects is already so large that isotropization of the sample actually occurs. Judging by our experimental results, the LPs, which determine the behavior of both the FLC and the PG in HTSC, perceive this state of the crystal as a sample "without defects." This conclusion seems quite reasonable, since it explains the temperature dependences of both the FLC and the PG obtained for the single crystal at $\varphi_5$. It can also be assumed that metal-insulator transition in the YBCO single crystal should apparently occur from a quasi-isotropic state. But, strictly speaking, the physics of all these processes is still not completely clear. Interestingly, in contrast to the rather peculiar $\Delta^*(T)$ dependences observed for single crystals irradiated with different $\varphi$, the value of $\Delta^*(T_G)$ decreases linearly with the irradiation doses (inset in Fig. 8), which corresponds well to the phase diagram proposed in [43] (see Fig. 28 in [43]). This allows us to conclude that the influence of irradiation defects on the FCPs, responsible for $\sigma'(T)$ near $T_c$, and on the LPs, determining PG, $\Delta^*(T)$, at higher temperatures, is significantly different. It is also worth noting that our results thus confirm the presence of paired fermions (local pairs) in HTSCs in the region of existence of the pseudogap.

In summary, to comprehend the obtained results, it is necessary to keep in mind that below $T^*$ in HTSC there are two types of charge carriers, namely, normal charge carriers and local pairs (LPs) [1–5,9–12], which, as we see, interact differently with irradiation. This assumption is confirmed by the absence of specific features for $\rho(\varphi)$ and $T_c(\varphi)$ (Fig. 2) and their presence on the dose dependencies $\sigma'(T)$ (Fig. 5) and $\Delta^*(T)$ (Fig. 8). Recall that the main result of the effect of electron irradiation on the crystal is the effective displacement of most of its atoms, which leads to the formation of numerous point and cluster defects that increase the resistance (Figs. 1–3). However, it is unlikely that defects directly affect the LPs, which are believed to carry electrical current without dissipation. Thus, it is important to understand that the $\rho(\varphi)$ dependence is determined by the behavior of normal charge carriers, which depends on the degree of disorder in the crystal. Whereas the $T_c(\varphi)$, $\sigma'(T)(\varphi)$, and $\Delta^*(T)(\varphi)$ dependencies are determined, as we have shown, by the behavior of LPs under irradiation. In turn, the behavior of LPs is determined by the AG pair-breaking mechanism of LPs at small $\varphi$ and the EK mechanism of quantum fluctuations of the order parameter at a higher degree of disorder [38]. Most likely, it is these two mechanisms that determine the linear dose dependence of $T_c$ (Fig. 2).

## IV. CONCLUSION

Using high-energy electron irradiation we succeeded in producing a fairly controlled disorder in untwined optimally doped YBCO single crystals by creating numerous uniformly distributed defects. This allowed for a detailed study of the temperature dependence of the resistivity and excess conductivity of a series of crystals irradiated with electrons of energy 2.5 MeV at doses from $\varphi_1 = 0$ to $\varphi_{12} = 23 \times 10^{19}$ e/cm². We found that the single crystals retained pseudogap for doses up to $\varphi_5 = 5.6 \times 10^{19}$ e/cm². For these crystals, the temperature dependences of the fluctuation conductivity (FLC) and the pseudogap (PG) derived from the excess conductivity were carefully analyzed in the frame of different theories, including our FLC approach.

A linear increase in resistivity followed by a linear decrease in $T_c$ and in the PG value, $\Delta^*(T_G)$, where $T_G$ is the Ginzburg temperature, was observed for all superconducting single crystals, i.e., those irradiated with electrons to the dose of $\varphi_5$. For relatively small $\varphi$, the linear $T_c(\varphi)$ dependence can be described by the Abrikosov-Gorkov (AG) theory of pair breaking, and for larger $\varphi$, by the Emery-Kivelson (EK) theory, which takes into account the suppression of $T_c$ by quantum phase fluctuations caused by irradiation defects. We found that as $\varphi$ increases, an AG-EK crossover appears in our untwined optimally doped YBCO single crystals. Interestingly, there are no specific features observed in $\rho(\varphi)$ and $T_c(\varphi)$ indicating this crossover. However, an unexpected increase in the PG opening temperature, $T^*$, and in the $\Delta^*(T)$ values is observed for the single crystal irradiated with $\varphi_3 = 2.5 \times 10^{19}$ e/cm² (see Figs. 3 and 8). This corresponds to the AG-EK crossover and suggests the possible decrease in DOS and very likely the transformation of the Fermi surface due to the irradiation defects introduced. Moreover, for $\varphi_3$ the distance between the CuO₂ conducting planes $d_{01}$, the coherence lengths $\xi_c(0)$ and $\xi_{ab}(0)$ and the range of superconducting fluctuations $\Delta T_{fl}$ increase rapidly (see Table III), and the contribution of 2D Maki-Thompson fluctuations surprisingly changes into the contribution of 2D Aslamazov-Larkin ones. With a further increase in $\varphi$, many parameters of the PG are greatly reduced and an unusual shape of $\Delta^*(T)$ is observed. One could expect that with an increase in the irradiation dose to $\varphi_5 = 5.6 \times 10^{19}$ e/cm² the PG would be strongly suppressed. However, very surprisingly, this did not happen. On the contrary, the FLC demonstrates the properties typical for well-structured YBCO and a very uniform $\Delta^*(T)$ dependence is observed, similar to that obtained for not irradiated or low-dose (up to $\varphi_3$) irradiated single crystals. This anomalous behavior of $\Delta^*(T)$ is in good agreement with the other results of our FLC analysis.

Thus, it was shown that the formation of various defects in YBCO single crystals under the influence of high-energy electron irradiation is a nonmonotonic process and has its own specificity. This specificity can be detected by analyzing the FLC and PG properties, which turned out to be much more sensitive to changes in internal electronic subsystems affected by induced defects. A generalization of the very peculiar features of the FLC and PG behavior observed in untwined YBCO single crystals under electron irradiation allows us to draw some more universal conclusions. The absence of any unusual features for $\rho(\varphi)$ and $T_c(\varphi)$ dependencies indicates that interaction of the irradiation defects with normal charge carriers, responsible for resistivity, and with local pairs, responsible for FLC and PG, differs significantly. Then, taking into account the diversity of the $\Delta^*(T)$ dependencies for different $\varphi$ and remembering the constant linear decrease in the $\Delta^*(T_G)$ values with $\varphi$, we can further conclude that the effect of defects on small tightly bound local pairs, responsible for the PG properties at high temperatures, and on large fluctuating Cooper pairs near $T_c$, responsible for the behavior of $\Delta^*(T_G)$, are also completely different. In addition, all findings confirm the conclusion we drew in our previous work that the more defects, the more isotropic the sample becomes [31]. The analysis of the results within our local pair model allows us not only to rationally explain the observed features, but also confirms the existence of paired fermions in the form of local pairs in HTSCs at $T \lesssim T^*$, the interaction of which with irradiation defects determines the unusual behavior of FLC and PG revealed in the experiment.

### ACKNOWLEDGMENT

A.L.S. thanks the Division of Low Temperatures and Superconductivity, INTiBS Wroclaw, Poland, for their hospitality.